\RequirePackage{ifpdf}
\ifpdf 
\documentclass[pdftex]{sigma}
\else
\documentclass{sigma}
\fi

\newtheorem{assumption}{Assumption}

\newcommand{\N}{\mathfrak N}

\newcommand{\mc}{\mathcal}

\newcommand{\D}{{\mc D}}

\newcommand{\F}{{\cal F}}

\newcommand{\Lc}{{\cal L}}

\newcommand{\Hil}{\mc H}

\begin{document}

\newcommand{\1}{1 \!\! 1}

\allowdisplaybreaks

\renewcommand{\thefootnote}{$\star$}

\renewcommand{\PaperNumber}{093}

\FirstPageHeading

\ShortArticleName{Pseudo-Bosons from Landau Levels}

\ArticleName{Pseudo-Bosons from Landau Levels\footnote{This
paper is a contribution to the Proceedings of the Workshop ``Supersymmetric Quantum Mechanics and Spectral Design'' (July 18--30, 2010, Benasque, Spain). The full collection
is available at
\href{http://www.emis.de/journals/SIGMA/SUSYQM2010.html}{http://www.emis.de/journals/SIGMA/SUSYQM2010.html}}}

\Author{Fabio BAGARELLO}

\AuthorNameForHeading{F.~Bagarello}

\Address{Dipartimento di Metodi e Modelli Matematici,
Facolt\`a di Ingegneria,\\ Universit\`a di Palermo, I-90128  Palermo, Italy}
\Email{\href{mailto:bagarell@unipa.it}{bagarell@unipa.it}}
\URLaddress{\url{http://www.unipa.it/~bagarell/}}

\ArticleDates{Received October 25, 2010, in f\/inal form December 02, 2010;  Published online December 12, 2010}

\Abstract{We construct examples of pseudo-bosons in two dimensions arising from the Hamiltonian for the Landau levels. We also prove a no-go result showing that non-linear combinations of bosonic creation and annihilation operators cannot give rise to pseudo-bosons.}


\Classification{81Q65; 65H17}

\renewcommand{\thefootnote}{\arabic{footnote}}
\setcounter{footnote}{0}

\section{Introduction}

In a series of recent papers \cite{bagpb1,bagpb2,bagpb3,bagcal,bagijtp,bagpb6}, we have investigated some mathematical aspects of the
so-called {\em pseudo-bosons}, originally introduced by Trifonov\footnote{It should be mentioned that pseudo-bosons already appeared in \cite{bes} but with a dif\/ferent meaning.}
in \cite{tri}. They arise from the canonical commutation relation
$[a,a^\dagger]=\1$ upon replacing $a^\dagger$ by another (unbounded)
opera\-tor~$b$ not (in general) related to~$a$: $[a,b]=\1$. We have
shown that, under suitable assumptions, $N=ba$ and $N^\dagger=a^\dagger b^\dagger$ can be both
diagonalized, and that their spectra coincide with the set of
natural numbers (including 0), ${\Bbb N}_0$. However the sets of
related eigenvectors are not orthonormal (o.n.) bases but,
nevertheless, they are automatically { biorthonormal\/}. In most of the
examples considered so far, they are bases of the Hilbert space of the system,
$\Hil$, and, in some cases, they turn out to be {\em Riesz bases\/}.

In \cite{bagpb4} and \cite{abg} some physical examples arising from  quantum mechanics have been discussed. In particular, these examples have suggested the introduction of a dif\/ference between what we have called {\em regular pseudo-bosons} and {\em pseudo-bosons}, to better focus on what we believe are the mathematical or on the physical aspects of these {\em particles}. Indeed all the examples of regular pseudo-bosons considered so far arise from Riesz bases~\cite{bagcal},  with a rather mathematical construction, while pseudo-bosons are those which one can f\/ind when starting with the Hamiltonian of some realistic quantum system.

In this paper, after a short review of the general framework, we discuss a two-dimensional example arising from the Hamiltonian of the Landau levels. It should be stressed that this example is of a completely dif\/ferent kind than those considered in~\cite{abg}, where a modif\/ied version of the Landau levels have been considered.

We close the paper with a no-go result, suggesting that non-linear combinations of ordinary bosonic creation and annihilation operators, even if they produce pseudo-bosonic commutation rules, cannot satisfy the Assumptions of our construction, see Section~\ref{section2}.

\section{The commutation rules}\label{section2}

In this section we will review a $d$-dimensional version  of
what originally proposed in \cite{bagpb1,bagpb6}.

Let $\Hil$ be a given Hilbert space with scalar product
$\langle \cdot ,\cdot\rangle$ and related norm $\|\cdot\|$. We introduce $d$
pairs of operators, $a_j$ and $b_j$, $j=1,2,\ldots,d$, acting on $\Hil$ and
satisfying the following commutation rules
\begin{gather}
 [a_j,b_k]=\delta_{j,k} \1,
\label{21}
\end{gather}  $j,k=1,2,\ldots,d$. Of course, these collapse to the CCR's for $d$
independent modes if $b_j=a^\dagger_j$, $j=1,2,\ldots,d$. It is well known
that $a_j$ and $b_j$ are unbounded operators, so they cannot be
def\/ined on all of $\Hil$. Following \cite{bagpb1}, and writing
$D^\infty(X):=\cap_{p\geq0}D(X^p)$ (the common  domain of all the powers of the
operator $X$), we consider the
following:

\begin{assumption}\label{Assumption1}
There exists a non-zero
$\varphi_{\bf 0}\in\Hil$ such that $a_j\varphi_{\bf 0}=0$, $j=1,2,\ldots,d$,
and $\varphi_{\bf 0}\in D^\infty(b_1)\cap D^\infty(b_2)\cap\cdots\cap D^\infty(b_d)$.
\end{assumption}

\begin{assumption}\label{Assumption2}
There exists a non-zero $\Psi_{\bf 0}\in\Hil$
such that $b_j^\dagger\Psi_{\bf 0}=0$, $j=1,2,\ldots,d$, and $\Psi_{\bf 0}\in
D^\infty(a_1^\dagger)\cap D^\infty(a_2^\dagger)\cap\cdots\cap D^\infty(a_d^\dagger)$.
\end{assumption}

Under these assumptions we can introduce the following vectors in
$\Hil$:
\begin{gather*}
\varphi_{\bf n}:=\varphi_{n_1,n_2,\ldots,n_d}=\frac{1}{\sqrt{n_1!n_2!\cdots n_d!}} b_1^{n_1} b_2^{n_2}\cdots b_d^{n_d} \varphi_{\bf 0},
\nonumber\\
\Psi_{\bf n}:=\Psi_{n_1,n_2,\ldots,n_d}=\frac{1}{\sqrt{n_1!n_2!\cdots n_d!}} {a_1^\dagger}^{n_1} {a_2^\dagger}^{n_2}\cdots {a_d^\dagger}^{n_d}\,\Psi_{\bf 0},
\end{gather*} $n_j=0, 1, 2,\ldots$ for all $j=1,2,\ldots,d$. Let us now def\/ine the unbounded
operators $N_j:=b_ja_j$ and $\N_j:=N_j^\dagger=a_j^\dagger
b_j^\dagger$, $j=1,2,\ldots,d$.  Each
$\varphi_{\bf n}$ belongs to the domain of $N_j$, $D(N_j)$, and
$\Psi_{\bf n}\in D(\N_j)$, for all possible $\bf n$. Moreover,
\begin{gather*}
 N_j\varphi_{\bf n}=n_j\varphi_{\bf n},  \qquad \N_j\Psi_{\bf n}=n_j\Psi_{\bf n}.
\end{gather*}

Under the above assumptions, and if we chose the normalization of
$\Psi_{\bf 0}$ and $\varphi_{\bf 0}$ in such a way that
$\langle \Psi_{\bf 0},\varphi_{\bf 0}\rangle =1$, we f\/ind that
\begin{gather*}
\langle \Psi_{\bf n},\varphi_{\bf m}\rangle =\delta_{\bf n,m}=\prod_{j=1}^d \delta_{n_j,m_j}. 
\end{gather*}
 This means that the sets
$\F_\Psi=\{\Psi_{\bf n}\}$ and
$\F_\varphi=\{\varphi_{\bf n}\}$ are {\em biorthonormal} and,
because of this, the vectors of each set are linearly independent.
If we now call $\D_\varphi$ and $\D_\Psi$ respectively the linear
span of  $\F_\varphi$ and $\F_\Psi$, and $\Hil_\varphi$ and
$\Hil_\Psi$ their closures, then
\begin{gather*}
f=\sum_{\bf n}
\langle \Psi_{\bf n},f\rangle \,\varphi_{\bf n}, \qquad \forall\,
f\in\Hil_\varphi,\qquad  h=\sum_{\bf n}
\langle \varphi_{\bf n},h\rangle \,\Psi_{\bf n}, \quad \forall\,
h\in\Hil_\Psi. 
\end{gather*}
 What is not in general ensured is
that
$\Hil_\varphi=\Hil_\Psi=\Hil$. Indeed, we can only state that
$\Hil_\varphi\subseteq\Hil$ and $\Hil_\Psi\subseteq\Hil$. However,
motivated by the examples discussed so far in the literature,  we consider

\begin{assumption} \label{Assumption3}
The above Hilbert spaces all coincide:
$\Hil_\varphi=\Hil_\Psi=\Hil$.
\end{assumption}

This means, in particular,
that both $\F_\varphi$ and $\F_\Psi$ are bases of $\Hil$. The resolution of the identity in the bra-ket formalism looks like
\[
\sum_{\bf n}
|\varphi_{\bf n}\rangle \langle \Psi_{\bf n}|=\sum_{\bf n}
|\Psi_{\bf n}\rangle \langle \varphi_{\bf n}|=\1.
\]
 Let us
now introduce the operators $S_\varphi$ and $S_\Psi$ via their
action respectively on  $\F_\Psi$ and $\F_\varphi$:
\begin{gather*}
S_\varphi\Psi_{\bf n}=\varphi_{\bf n},\qquad
S_\Psi\varphi_{\bf n}=\Psi_{\bf n}, 
\end{gather*}
for all $\bf n$, which also imply that
$\Psi_{\bf n}=(S_\Psi\,S_\varphi)\Psi_{\bf n}$ and
$\varphi_{\bf n}=(S_\varphi \,S_\Psi)\varphi_{\bf n}$, for all
$\bf n$. Hence
\begin{gather*}
 S_\Psi S_\varphi=S_\varphi S_\Psi=\1 \quad
\Rightarrow \quad S_\Psi=S_\varphi^{-1}. 
\end{gather*}
In other
words, both $S_\Psi$ and $S_\varphi$ are invertible and one is the
inverse of the other. Furthermore, we can also check that they are
both positive, well def\/ined and symmetric~\cite{bagpb1}. Moreover, it is possible to write these operators  as
\begin{gather*}
S_\varphi=\sum_{\bf n} |\varphi_{\bf n}\rangle \langle \varphi_{\bf n}|,\qquad S_\Psi=\sum_{\bf n}
 |\Psi_{\bf n}\rangle \langle \Psi_{\bf n}|. 
 \end{gather*}
 These expressions are
only formal, at this stage, since the series may not converge in
the uniform topology and the operators $S_\varphi$ and $S_\Psi$ could be unbounded.
Indeed we know~\cite{you}, that two biorthonormal bases are related by a bounded operator, with bounded inverse, if and only if they are Riesz bases\footnote{Recall that a set of vectors $\phi_1, \phi_2 , \phi_3 ,   \ldots   $, is a Riesz basis of a Hilbert space $\mathcal H$, if there exists a bounded operator $V$, with bounded inverse, on $\mathcal H$, and an o.n.\ basis of $\Hil$,  $\varphi_1, \varphi_2 , \varphi_3 ,   \ldots   $, such that $\phi_j=V\varphi_j$, for all $j=1, 2, 3,\ldots$}. This is why in \cite{bagpb1} we have also considerered

\begin{assumption}\label{Assumption4}
$\F_\varphi$ and $\F_\Psi$ are  both Riesz bases.
\end{assumption}

Therefore, as already stated, $S_\varphi$ and $S_\Psi$ are bounded operators and their domains  can be taken to be all of $\Hil$. While Assumptions~\ref{Assumption1},~\ref{Assumption2} and~\ref{Assumption3} are quite often satisf\/ied, \cite{bagrev},  it is quite dif\/f\/icult to f\/ind {\it physical} examples satisfying also Assumption~\ref{Assumption4}. On the other hand, it is rather easy to f\/ind {\it mathematical} examples satisfying all the assumptions, see~\cite{bagpb1,bagpb6}. This is why in~\cite{bagpb4} we have introduced a dif\/ference in the notation: we have called {\em pseudo-bosons} (PB) those satisfying  the f\/irst three assumptions, while, if they also satisfy Assumption~\ref{Assumption4}, they are called {\em regular pseudo-bosons} (RPB).

As already discussed in our previous papers,  these
$d$-dimensional pseudo-bosons give rise to interesting
intertwining relations among non self-adjoint operators, see in particular~\cite{bagpb3} and references therein. For instance, it is easy to
check that
\begin{gather*}
 S_\Psi N_j=\N_jS_\Psi \qquad \mbox{and}\qquad
N_j S_\varphi=S_\varphi \N_j, 
\end{gather*}
$j=1,2,\ldots,d$. This is
related to the fact that the spectra of, say, $N_1$ and $\N_1$,
coincide and that their eigenvectors are related by the operators
$S_\varphi$ and $S_\Psi$, in agreement with the literature on
intertwining operators~\cite{intop,bag1}.

\section{The example}\label{section3}

In this section we will consider an example arising from a quantum mechanical system, i.e.
 a single electron moving on a two-dimensional
plane and subject to a uniform magnetic f\/ield along the
$z$-direction. Taking $\hbar=m=\frac{eB}{c}=1$, the Hamiltonian of the electron is given by the operator
\begin{gather}
H_1={\frac 12} \left(\underline p-\underline A(r)\right)^2={\frac
12}  \left(p_x+{\frac y2}\right)^2+{\frac 12} \left(p_y-{\frac
x2} \right)^2, \label{31}
\end{gather}
where we have used minimal coupling and the symmetric gauge $\vec
A=\frac{1}{2}(-y,x,0)$. The Hilbert space of the system is $\Hil=\Lc^2({\Bbb R}^2)$.

The spectrum of this Hamiltonian is easily obtained by f\/irst
introducing the new variables
  \begin{gather}
\label{32}
  Q_1= p_x+y/2, \qquad     P_1= p_y-x/2.
\end{gather}
In terms of $P_1$ and $Q_1$ the single electron Hamiltonian,
$H_1$, can be rewritten as
 \begin{gather*}
  H_1=\frac{1}{2}(Q_1^2 + P_1^2).
\end{gather*}
The transformation (\ref{32}) is part of a canonical map from the
 variables $(x,y,p_x,p_y)$ to $(Q_1,Q_2,P_1,$ $P_2)$,
where
 \begin{gather*}
   Q_2= p_y+x/2, \qquad
  P_2= p_x-y/2,
\end{gather*}
which can be used to construct a second Hamiltonian $ H_2=\frac{1}{2}(Q_2^2 + P_2^2)$. Since
$ [x, p_x] = [y, p_y] = i$, $[x,p_y] = [y,p_x] = [x,y] = [p_x , p_y ] = 0,$
we deduce that
\begin{gather*}
 [Q_1,P_1] = [Q_2,P_2]=i, \qquad  [Q_1,P_2]=[Q_2,P_1]=[Q_1,Q_2]=[P_1,P_2]=0,
\end{gather*}
so that $[H_1,H_2]=0$. The two Hamiltonians correspond to two opposite magnetic f\/ields, respectively along $+\hat k$ and $-\hat k$. Let us now introduce the operators
\begin{gather*}
A_k=\frac{1}{\sqrt{2}}\left(Q_k+iP_k\right),
\end{gather*}
$k=1,2$, together with their adjoints. Then $[A_k,A_l^\dagger]=\delta_{k,l}\1$, the other commutators being zero. In terms of these operators we can write $H_k=A_k^\dagger A_k+\frac{1}{2} \1$, $k=1,2$, whose eigenvectors are $\Phi^{(k)}_n=\frac{1}{\sqrt{n!}} (A_k^\dagger)^n\Phi^{(k)}_0$, where $k=1,2$, $n=0,1,2,\ldots$ and $\Phi^{(k)}_0$ is the vacuum of $A_k$: $A_k\Phi^{(k)}_0=0$. Furthermore we have $\langle \Phi^{(k)}_n,\Phi^{(k)}_m\rangle =\delta_{n,m}$ and $H_k\Phi^{(k)}_n=\left(n+\frac{1}{2}\right)\Phi^{(k)}_n$, for $k=1,2$. It is natural to introduce the  sets $\F_k:=\big\{\Phi^{(k)}_n,\,n\geq0\big\}$, $k=1,2$, and the closures of their linear span, $\Hil_1$ and $\Hil_2$. Hence, by construction, $\F_k$ is an o.n. basis of $\Hil_k$. Moreover, we can also introduce an o.n.\ basis of $\Hil$ as the set $\F_\Phi$ whose vectors are def\/ined as follows:
\begin{gather*}
\Phi_{n,m}:=\frac{1}{\sqrt{n! m! }}(A_1^\dagger)^n(A_2^\dagger)^m \Phi_{0,0},
\end{gather*}
where $\Phi_{0,0}:=\Phi^{(1)}_0\otimes\Phi^{(2)}_0$ is such that $A_1\Phi_{0,0}=A_2\Phi_{0,0}=0$. It is clear that $\Phi_{n,m}=\Phi^{(1)}_n\otimes\Phi^{(2)}_m$ and that $\Hil=\Hil_1\otimes\Hil_2$.

\subsection[Pseudo-bosons in ${\mathcal H}_1$]{Pseudo-bosons in $\boldsymbol{{\mathcal H}_1}$}\label{section3.1}

Let us now def\/ine the following operators: $A_1(\alpha)=A_1$ and $B_1(\alpha)=A_1^\dagger+2\alpha A_1$, where $\alpha$ is a~f\/ixed complex number. It is clear that, for $\alpha\neq0$, $A_1(\alpha)^\dagger\neq B_1(\alpha)$. Moreover, $[A_1(\alpha),B_1(\alpha)]=\1$, $\forall\, \alpha$. Hence, we recover (\ref{21}) for $d=1$ in $\Hil_1$. We want to show that $A_1(\alpha)$ and $B_1(\alpha)$ generate PB in $\Hil_1$ which are not regular.

To begin with, we def\/ine $\varphi_0^{(1)}(\alpha):=\Phi_0^{(1)}$. This non zero vector of $\Hil_1$ satisf\/ies Assumption~\ref{Assumption1}: $A_1(\alpha)\varphi_0^{(1)}(\alpha)=0$, clearly, and $\varphi_0^{(1)}(\alpha)\in D^\infty(B_1(\alpha))$. This follows from the fact that, since $B_1(\alpha)=A_1^\dagger+2\alpha A_1$, $B_1(\alpha)^n\varphi_0^{(1)}(\alpha)$ is a f\/inite linear combination of the vectors $\Phi_0^{(1)}, \Phi_1^{(1)}, \ldots$, $\Phi_n^{(1)}$, which is clearly a vector of $\Hil_1$.

Before considering Assumption~\ref{Assumption2}, it is convenient to observe that, introducing the following invertible and densely def\/ined operator $U_1(\alpha):=e^{\alpha A_1^2}$, we can write
\begin{gather}
A_1(\alpha)=U_1(\alpha)A_1U_1(\alpha)^{-1},\qquad  B_1(\alpha)=U_1(\alpha)A_1^\dagger U_1(\alpha)^{-1}, \nonumber\\
\varphi_n^{(1)}(\alpha):=\frac{1}{\sqrt{n!}}B_1(\alpha)^n\varphi_0^{(1)}(\alpha)=U_1(\alpha)\Phi_n^{(1)},
\label{38}
\end{gather}
for all $n\geq0$. Of course, $\varphi_n^{(1)}(\alpha)$ is well def\/ined for all $n\geq0$ since, as we have seen, $B_1(\alpha)^n\varphi_0^{(1)}(\alpha)$ is well def\/ined for all complex $\alpha$. Now, if we def\/ine (at least formally, at this stage)
\begin{gather}
\Psi_0^{(1)}(\alpha):=\big(U_1(\alpha)^\dagger\big)^{-1}\Phi_0^{(1)},
\label{39}
\end{gather}
it is possible to show that, if $|\alpha|<\frac{1}{2}$: (i)~$\Psi_0^{(1)}(\alpha)$ is well def\/ined in $\Hil_1$, and is dif\/ferent from zero; (ii)~$B_1(\alpha)^\dagger\Psi_0^{(1)}(\alpha)=0$; (iii)~$\Psi_0^{(1)}(\alpha)\in D^\infty(A_1^\dagger)$. It is furthermore possible to check that, for the same values of $\alpha$,
\begin{gather}
\Psi_n^{(1)}(\alpha):=\frac{1}{\sqrt{n!}}\big(A_1(\alpha)^\dagger\big)^n\Psi_0^{(1)}(\alpha)=
\big(U_1(\alpha)^\dagger\big)^{-1}\Phi_n^{(1)}.
\label{310}
\end{gather}
Let us prove point (iii) above. We have, for all $n\geq0$,
\[
\big(A_1^\dagger\big)^n e^{-\overline{\alpha} {A_1^\dagger}^2}\Phi_0^{(1)}=\sum_{k=0}^\infty\frac{(-\overline{\alpha})^k}{k!}
 \sqrt{(2k+n)!} \Phi_{2k+n}^{(1)},
\]
so that
\[
\big\|(A_1^\dagger)^n e^{-\overline{\alpha} {A_1^\dagger}^2}\Phi_0^{(1)}\big\|^2=
\sum_{k=0}^\infty\frac{|\alpha|^{2k}}{(k!)^2} (2k+n)!,
\]
which converges inside the disk $|\alpha|<\frac{1}{2}$. In particular, if $n=0$, this implies the statement in~(i) above. The proof of~(ii) is trivial and the last equality in~(\ref{310}) can be deduced  using~(\ref{38}) and~(\ref{39}) in the def\/inition $\Psi_n^{(1)}(\alpha):=\frac{1}{\sqrt{n!}}(A_1(\alpha)^\dagger)^n\Psi_0^{(1)}(\alpha)$. This, as we have seen, is well def\/ined if  $|\alpha|<\frac{1}{2}$, while, for $|\alpha|>\frac{1}{2}$ all the procedure makes no sense, since the vectors we are using do not belong to the Hilbert space. Now, biorthonormality of the two sets $\F_{\varphi^{(1)}}:=\{\varphi_n^{(1)}(\alpha), \,n\geq0\}$ and $\F_{\Psi^{(1)}}:=\{\Psi_n^{(1)}(\alpha), \,n\geq0\}$ follows directly from their def\/initions:
\begin{gather*}
\langle \varphi_n^{(1)}(\alpha),\Psi_m^{(1)}(\alpha)\rangle =\langle U_1(\alpha)\Phi_n^{(1)},
\big(U_1(\alpha)^\dagger\big)^{-1}\Phi_m^{(1)}\rangle =\langle \Phi_n^{(1)},
\Phi_m^{(1)}\rangle =\delta_{n,m}.
\end{gather*}
The proof of Assumption~\ref{Assumption3} goes as follows:

First of all, as we have already stated, it is possible to check that for all $n\geq0$ we have $\varphi_n^{(1)}(\alpha)=\Phi_n^{(1)}+\sum_{k=0}^{n-1}d_k\Phi_k^{(1)}$, for some constants $\{d_k, \,k=0,1,\ldots,n-1\}$.

Secondly, using induction on $n$ and this simple remark we can prove that, {\em if $f\in\Hil_1$ is such that $\langle f,\varphi_k^{(1)}(\alpha)\rangle =0$ for $k=0,1,\ldots,n$, then $\langle f,\Phi_k^{(1)}\rangle =0$ for $k=0,1,\ldots,n$ as well.} Therefore, if $f$ is orthogonal to all the $\varphi_k^{(1)}(\alpha)$'s, it is also orthogonal to all the $\Phi_k^{(1)}$'s, whose set is complete in $\Hil_1$. Hence $f=0$, so that $\F_{\varphi^{(1)}}$ is also complete in $\Hil_1$.

As a consequence, being the vectors of $\F_{\varphi^{(1)}}$ linearly independent and complete in $\Hil_1$, they are a basis of $\Hil_1$. In particular we f\/ind that, for all $f\in\Hil_1$, the following expansion holds true: $f=\sum_{k=0}^\infty \langle \Psi_n^{(1)}(\alpha),f\rangle \varphi_n^{(1)}(\alpha)$. Then, for all $f,g\in\Hil_1$,
\[
\langle g,f\rangle =\Big\langle g,\sum_{k=0}^\infty \langle \Psi_n^{(1)}(\alpha),f\rangle \varphi_n^{(1)}(\alpha)\Big\rangle =
\Big\langle \sum_{k=0}^\infty \langle \varphi_n^{(1)}(\alpha),g\rangle \Psi_n^{(1)}(\alpha),f\Big\rangle ,
\]
which, since $f$ could be any vector in $\Hil_1$, implies that  $g=\sum_{k=0}^\infty \langle \varphi_n^{(1)}(\alpha),g\rangle \Psi_n^{(1)}(\alpha)$: $\F_{\Psi^{(1)}}$ is a~basis of $\Hil_1$ as well, and Assumption~\ref{Assumption3} is satisf\/ied. Finally, Assumption~\ref{Assumption4} is not satisf\/ied since, for instance, the operator $\left(U_1(\alpha)^\dagger\right)^{-1}$ is unbounded~\cite{you}.

\begin{remark} It might be worth stressing that, while it is quite easy to check that the set $\F_{\varphi^{(1)}}$ is complete in $D(U(\alpha)^\dagger)$, it is not trivial at all to check that it is also complete in $\Hil_1$. This is the reason why we have used the above procedure.
\end{remark}

It is not hard to deduce the expression of two non self-adjoint operators which admit $\varphi_n^{(1)}(\alpha)$ and $\Psi_n^{(1)}(\alpha)$ as eigenstates. For that we def\/ine f\/irst $h_1(\alpha):=U_1(\alpha)H_1U_1(\alpha)^{-1}=B_1(\alpha)A_1(\alpha)+\frac{1}{2} \1$, which, in coordinate representation, looks like
\[
h_1(\alpha)=\left(\frac{1}{2}+\alpha\right)\left(p_x+\frac{y}{2}\right)^2+
\left(\frac{1}{2}-\alpha\right)\left(p_y-\frac{x}{2}\right)^2+2i\alpha\left(p_x+\frac{y}{2}\right)
\left(p_y-\frac{x}{2}\right)+\alpha\1.
\]
We can also introduce $h_1(\alpha)^\dagger$, which is clearly dif\/ferent from $h_1(\alpha)$. Now, as expected from general facts in the theory of intertwining operators~\cite{intop}, we see that
\begin{gather*}
h_1(\alpha)\varphi_n^{(1)}(\alpha)=(n+1/2)\varphi_n^{(1)}(\alpha),\qquad h_1(\alpha)^\dagger\Psi_n^{(1)}(\alpha)=(n+1/2)\Psi_n^{(1)}(\alpha),
\end{gather*}
for all $n\geq0$.

\subsection[Pseudo-bosons in ${\mathcal H}_2$]{Pseudo-bosons in $\boldsymbol{{\mathcal H}_2}$}\label{section3.2}

In this subsection we will consider an analogous construction in $\Hil_2$, i.e.\ in the Hilbert space related to the uniform magnetic f\/ield along $-\hat k$. To make the situation more interesting, and to avoid repeating essentially the same procedure  considered above, instead of introducing an operator like $e^{\beta A_2^2}$ we consider
\begin{gather*}
U_2(\beta):=e^{\beta {A_2^\dagger}^2},
\end{gather*}
with $\beta\in\Bbb{C}$. Then we def\/ine
\begin{gather}
A_2(\beta):=U_2(\beta)A_2U_2(\beta)^{-1}=A_2-2\beta A_2^\dagger,\nonumber\\
B_2(\beta):=U_2(\beta)A_2^\dagger U_2(\beta)^{-1}=A_2^\dagger.
\label{314}
\end{gather}
These are pseudo-bosonic operators in $\Hil_2$: $[A_2(\beta),B_2(\beta)]=\1$, and $A_2(\beta)^\dagger\neq B_2(\beta)$, for $\beta\neq0$. Then, once again, it may be interesting to consider Assumptions~\ref{Assumption1}--\ref{Assumption4}.

If $|\beta|<\frac{1}{2}$ Assumption~\ref{Assumption1} is satisf\/ied: let us def\/ine (formally, for the moment) $\varphi_0^{(2)}(\beta)=U_2(\beta)\Phi_0^{(2)}$. Then $A_2(\beta)\varphi_0^{(2)}(\beta)=0$. Moreover, since $[B_2(\beta),U_2(\beta)]=0$, $B_2(\beta)^n\varphi_0^{(2)}(\beta)=U_2(\beta){A_2^\dagger}^n\,\Phi_0^{(2)} $, which implies in particular that
\begin{gather*}
\varphi_n^{(2)}(\beta):=\frac{1}{\sqrt{n!}}B_2(\beta)^n\varphi_0^{(2)}(\beta)=U_2(\beta)\Phi_n^{(2)}.
\end{gather*}
Of course we have now to check that $\varphi_n^{(2)}(\beta)$ is a well def\/ined vector of $\Hil_2$ for all $n\geq0$. This would make the above formal def\/inition rigorous. The computation of $\|U_2(\beta)\Phi_n^{(2)}\|$ follows the same steps as that for $\|{U_1(\alpha)^\dagger}^{-1}\Phi_n^{(1)}\|$ of the previous section, and we get the same conclusion: the power series obtained for $\|U_2(\beta)\Phi_n^{(2)}\|^2$ converges if $|\beta|<\frac{1}{2}$, so that $\Phi_n^{(2)}\in D(U_2(\beta))$ for all $n\geq0$, inside this disk.

As for Assumption~\ref{Assumption2}, this is also satisf\/ied: to prove this it is enough to take $\Psi_0^{(2)}(\beta)=\Phi_0^{(2)}$. Then $B_2(\beta)^\dagger
\Psi_0^{(2)}(\beta)=A_2\Phi_0^{(2)}=0$. Also, since $U_2(\beta)^\dagger\Phi_0^{(2)}=\Phi_0^{(2)}$, formula (\ref{314}) implies that $\frac{1}{\sqrt{n!}}(A_2^\dagger(\beta))^n\Psi_0^{(2)}(\beta)=e^{-\overline{\beta}\,A_2^2}\Phi_n^{(2)}$, which is clearly a vector in $\Hil_2$ since it is a f\/inite linear combination of $\Phi_0^{(2)}, \Phi_1^{(2)},\ldots,\Phi_n^{(2)}$. This means that the vectors
\begin{gather*}
\Psi_n^{(2)}(\beta):=\frac{1}{\sqrt{n!}}{A_2(\beta)^\dagger}^n\Psi_0^{(2)}(\beta)=\big(U_2(\beta)^\dagger
\big)^{-1}\Phi_n^{(2)}
\end{gather*}
are well def\/ined in $\Hil_2$ for all~$n$, independently of~$\beta$. Once again we deduce that the vectors constructed here are biorthonormal,
\begin{gather*}
\langle \varphi_n^{(2)}(\beta),\Psi_m^{(2)}(\beta)\rangle =\delta_{n,m},
\end{gather*}
 and that they are eigenstates of two operators which are the adjoint one of the other, and which are related to $H_2$ by a similarity transformation:
\begin{gather*}
h_2(\beta):=U_2(\beta)H_2U_2(\beta)^{-1}= B_2(\beta)A_2(\beta)+\frac{1}{2}\1,
\end{gather*}
which in coordinate representation looks like
\[
h_2(\beta)=\left(\frac{1}{2}-\beta\right)\left(p_y+\frac{x}{2}\right)^2+
\left(\frac{1}{2}+\beta\right)\left(p_x-\frac{y}{2}\right)^2+2i\beta\left(p_y+\frac{x}{2}\right)
\left(p_x-\frac{y}{2}\right)+\beta\1.
\]
In particular we f\/ind that
\begin{gather*}
h_2(\beta)\varphi_n^{(2)}(\beta)=(n+1/2)\varphi_n^{(2)}(\beta),\qquad h_2(\beta)^\dagger\Psi_n^{(2)}(\beta)=(n+1/2)\Psi_n^{(2)}(\beta),
\end{gather*}
for all $n\geq0$. The same arguments used previously prove that $\F_{\varphi^{(2)}}:=\{\varphi_n^{(2)}(\beta),\,n\geq0\}$ and $\F_{\Psi^{(2)}}:=\{\Psi_n^{(2)}(\beta),\,n\geq0\}$ are both complete in~$\Hil_2$. More than this: they are biorthonormal bases but not Riesz bases.

\subsection[Pseudo-bosons in ${\mathcal H}$]{Pseudo-bosons in $\boldsymbol{{\mathcal H}}$}\label{section3.3}

We begin this section with the following remark: none of the above sets of functions is complete in $\Hil$. Hence we could try to f\/ind a dif\/ferent set of vectors, also labeled by a single quantum number, which is complete in $\Hil$. It is not hard to check that this is not possible, in general. Let us introduce, for instance, the following pseudo-bosonic operators: $X_{\alpha,\beta}:=\frac{1}{\sqrt{2}}\left(A_1(\alpha)+A_2(\beta)\right)$ and $Y_{\alpha,\beta}:=\frac{1}{\sqrt{2}}\left(B_1(\alpha)+B_2(\beta)\right)$. Then $[X_{\alpha,\beta},Y_{\alpha,\beta}]=\1$, $X_{\alpha,\beta}\neq Y_{\alpha,\beta}^\dagger$ in general and the vectors $\varphi_{0,0}(\alpha,\beta):=\varphi_{0}^{(1)}(\alpha)\otimes\varphi_{0}^{(2)}(\beta)$ and $\Psi_{0,0}(\alpha,\beta):=\Psi_{0}^{(1)}(\alpha)\otimes\Psi_{0}^{(2)}(\beta)$ satisfy Assumptions~\ref{Assumption1} and~\ref{Assumption2} of Section~\ref{section2}. However, it is not hard to check that the vectors $\eta_n(\alpha,\beta):=\frac{1}{\sqrt{n!}}Y_{\alpha,\beta}^n\varphi_{0,0}(\alpha,\beta)$, $n\geq0$, are not complete in $\Hil$: for that it is enough to consider the non zero vector $f=\Psi_{1}^{(1)}(\alpha)\otimes\Psi_{0}^{(2)}(\beta)-\Psi_{0}^{(1)}(\alpha)\otimes\Psi_{1}^{(2)}(\beta)$, which is non zero and orthogonal to all the $\eta_n(\alpha,\beta)$'s.

This is not surprising: in Section~\ref{section2}, in fact, we have proposed a dif\/ferent way to produce two biorthonormal bases of $\Hil$, in dimension larger than~1. For instance, in $d=2$ we expect that the vectors of these bases depend on two quantum numbers rather than just one. So we may proceed as follows: let $T(\alpha,\beta)$ be the following unbounded operator:
\begin{gather*}
T(\alpha,\beta):=U_1(\alpha)U_2(\beta)=e^{\alpha A_1^2+\beta {A_2^\dagger}^2}.
\end{gather*}
Then the vectors $\varphi_{0,0}(\alpha,\beta)$ and $\Psi_{0,0}(\alpha,\beta)$ introduced above can be def\/ined as $\varphi_{0,0}(\alpha,\beta)=T(\alpha,\beta)\Phi_{0,0}$ and $\Psi_{0,0}(\alpha,\beta)=\left(T(\alpha,\beta)^\dagger\right)^{-1}\Phi_{0,0}$. For what we have seen in the previous sections, these two vectors satisfy Assumptions~\ref{Assumption1} and~\ref{Assumption2}: $A_1(\alpha)\varphi_{0,0}(\alpha,\beta)= A_2(\beta)\varphi_{0,0}(\alpha,\beta)=0$, $B_1(\alpha)^\dagger\Psi_{0,0}(\alpha,\beta)= B_2(\beta)^\dagger\Psi_{0,0}(\alpha,\beta)=0$, and $\varphi_{0,0}(\alpha,\beta)\in D^\infty(B_1(\alpha))\cap D^\infty(B_2(\beta))$, $\Psi_{0,0}(\alpha,\beta)\in D^\infty(A_1(\alpha)^\dagger)\cap D^\infty(A_2(\beta)^\dagger)$.

Furthermore,  since $\Hil=\Hil_1\otimes\Hil_2$, the sets  $\F_{\varphi}\!:=\!\{\varphi_{n,m}(\alpha,\beta)\!:=\!\frac{1}{\sqrt{n!m!}}\!B_1(\alpha)^nB_2(\beta)^m
\varphi_{0,0}(\alpha,\beta)\}$ and  $\F_{\Psi}:=\{\Psi_{n,m}(\alpha,\beta):=\frac{1}{\sqrt{n!m!}}{A_1(\alpha)^\dagger}^n{A_2(\beta)^\dagger}^m
\Psi_{0,0}(\alpha,\beta)\}$ are complete in $\Hil$, so that Assumption~\ref{Assumption3} is also satisf\/ied.  Finally, Assumption~\ref{Assumption4} is not verif\/ied, so that we have found PB which are not regular. This is because $T(\alpha,\beta)$ is unbounded and since we can write $\varphi_{n,m}(\alpha,\beta)=T(\alpha,\beta)\Phi_{n,m}$ and $\Psi_{n,m}(\alpha,\beta)=\left(T(\alpha,\beta)^\dagger\right)^{-1}\Phi_{n,m}$, for all $n$ and $m$, \cite{you}.

\begin{remark}
The procedure outlined in this section clearly applies to any pair of uncoupled harmonic oscillators $h_1=a_1^\dagger a_1$ and $h_2=a_2^\dagger a_2$, $[a_i,a_j^\dagger]=\delta_{i,j}\,\1$, $i,j=1,2$, changing properly the def\/initions of the operators involved.
\end{remark}

\begin{remark}
Bi-coherent states like those in \cite{bagpb1} can be easily constructed from the ones for $A_1$ and $A_2$ using the operators $U_1(\alpha)$ and $U_2(\beta)$.
\end{remark}

\section{A no-go result}\label{section4}

We devote this short section to prove the following general no-go result: suppose $a$ and $a^\dagger$ are two operators acting on $\Hil$ and satisfying $[a,a^\dagger]=\1$. Then, for all $\alpha\neq0$, the operators $A:=a-\alpha {a^\dagger}^2$ and $B:=a^\dagger$ are such that $[A,B]=\1$, $A^\dagger\neq B$, but they do not satisfy Assumption~\ref{Assumption1}.

In fact, if such a non zero vector $\varphi_0\in\Hil$ exists, then it could be expanded in terms of the eigenvectors $\Phi_n:=\frac{{a^\dagger}^n}{\sqrt{n!}}\Phi_0$, $a\Phi_0=0$, of the number operator $N=a^\dagger a$: $\varphi_0=\sum_{n=0}^\infty c_n\Phi_n$, for some sequence $\{c_n, n\geq0\}$ such that $\sum_{n=0}^\infty|c_n|^2<\infty$. Condition $A\varphi_0=0$ can be rewritten as $a\varphi_0=\alpha {a^\dagger}^2\varphi_0$. Now, inserting in both sides of this equality the expansion for $\varphi_0$, and recalling  that $a^\dagger\Phi_n=\sqrt{n+1}\Phi_{n+1}$ and $a\Phi_n=\sqrt{n}\Phi_{n-1}$, $n\geq0$, we deduce the following relations between the coef\/f\/icients $c_n$: $c_1=c_2=0$ and $c_{n+1}\sqrt{n+1}=\alpha c_{n-2}\sqrt{(n-1)n}$, for all $n\geq2$. The solution of this recurrence relation is the following:
\[
c_3=\alpha c_0\frac{\sqrt{3!}}{3},\qquad c_6=\alpha^2 c_0\frac{\sqrt{6!}}{3\cdot6}, \qquad c_9=\alpha^3 c_0\frac{\sqrt{9!}}{3\cdot6\cdot9}, \qquad c_{12}=\alpha^4 c_0\frac{\sqrt{12!}}{3\cdot6\cdot9\cdot12},
\]
and so on. Then
\[
\varphi_0=c_0\left(\Phi_0+\sum_{k=1}^\infty\,\alpha^k\frac{\sqrt{(3k)!}}{1\cdot3\cdots3k}\,\Phi_{3k}\right).
\]
However, computing $\|\varphi_0\|$ we deduce that this series only converge if $\alpha=0$, i.e.\ if $A$ coincides with $a$ and $B$ with $a^\dagger$.

A similar results can be obtained considering the operators $A:=a-\alpha {a^\dagger}^n$ and $B:=a^\dagger-\beta\1$, $n\geq2$, $\alpha, \beta\in \Bbb{C}$. Again we f\/ind $[A,B]=\1$, $A^\dagger\neq B$, and again, with similar techniques, we deduce that they do not satisfy Assumption~\ref{Assumption1}. In the same way, if we def\/ine $A:=a-\alpha \1$ and $B:=a^\dagger-\beta a^m$, $m\geq2$, $\alpha, \beta\in \Bbb{C}$, we f\/ind that, in general, $[A,B]=\1$, $A^\dagger\neq B$, but they do not satisfy Assumption~\ref{Assumption2}. This  suggests that  if we try to def\/ine, starting from $a$ and $a^\dagger$, new operators $A=a+f(a,a^\dagger)$ and $B=a^\dagger+g(a,a^\dagger)$, only very special choices of $f$ and $g$ are compatible with the pseudo-bosonic structure.

\section{Conclusions}

We have seen how a non trivial example of two-dimensional PB arises from the Hamiltonian of the Landau levels. We want to stress once again that this is deeply dif\/ferent from what we have done in \cite{abg}, where the starting point was a generalized Hamiltonian obtained with a {\em smart} extension of that in~(\ref{31}) with the introduction of two related {\em superpotentials}. Among the other dif\/ferences, while the procedure outlined here works in any Hilbert space, the one in~\cite{abg} works only in~$\Lc^2({\Bbb R}^2)$. The fact that both here and in~\cite{abg} we get pseudo-bosons which are not regular is still another indication of the mathematical nature of RPB.

It is not dif\/f\/icult to modify or to generalize the results in Section~\ref{section3}, for instance changing the role of the operators $U_1$ and $U_2$, or modifying a bit their def\/initions. Maybe more interesting is to try to extend the no-go result of Section~\ref{section4} to other possible combinations of $a$ and $a^\dagger$: this is part of our work in progress.

\subsection*{Acknowledgements}
   The author would like to thank A.~Andrianov for his kind invitation and the local people in Benasque for their warm welcome.

\pdfbookmark[1]{References}{ref}
\LastPageEnding

\end{document}